\documentclass[11pt,a4paper,british,american]{article}
\usepackage{lmodern}

\usepackage[T1]{fontenc}
\usepackage[latin9]{inputenc}
\usepackage{amsmath}
\usepackage{amssymb}
\usepackage{esint}

\makeatletter

\pdfpageheight\paperheight
\pdfpagewidth\paperwidth

\usepackage{jcappub}
\numberwithin{equation}{section}

\pdfoutput=1
\renewcommand\[{\begin{equation}}
\renewcommand\]{\end{equation}}

\AtBeginDocument{
  
}

\makeatother

\usepackage{babel}
\begin{document}

\title{Many Faces of Mimetic Gravity }

\author[\selectlanguage{british}%
a\selectlanguage{american}%
]{Katrin Hammer\foreignlanguage{british}{}}

\author[\selectlanguage{british}%
b\selectlanguage{american}%
]{and Alexander Vikman\foreignlanguage{british}{}}

\selectlanguage{british}%

\affiliation[\selectlanguage{british}%
a\selectlanguage{american}%
]{\foreignlanguage{american}{Arnold Sommerfeld Center for Theoretical
Physics, }\\
\foreignlanguage{american}{Ludwig Maximilian University Munich, }\\
\foreignlanguage{american}{Theresienstr. 37, D-80333, Munich, Germany}}

\affiliation[\selectlanguage{british}%
b\selectlanguage{american}%
]{\foreignlanguage{american}{ Institute of Physics, the Academy of
Sciences of the Czech Republic, }\\
\foreignlanguage{american}{Na Slovance 2, CS-18221 Prague 8, Czech
Republic}\\
\foreignlanguage{american}{ and }\\
\foreignlanguage{american}{Yukawa Institute for Theoretical Physics,
Kyoto University, Kyoto 606-8502, Japan}}

\selectlanguage{american}%

\emailAdd{katrin.a.hammer@physik.uni-muenchen.de, vikman@fzu.cz}

\abstract{We consider the recently introduced mimetic gravity, which is a Weyl-symmetric
extension of the General Relativity and which can play a role of an
imperfect fluid-like Dark Matter with a small sound speed. In this
paper we discuss in details how this higher-derivative scalar-tensor
theory goes beyond the construction by Horndeski, keeping only one
scalar degree of freedom on top of two standard graviton polarizations.
In particular, we consider representations of the theory in different
sets of Weyl-invariant variables and connect this framework to the
singular Brans-Dicke theory. Further, we find solution of equations
of motion for the mimetic gravity in the synchronous reference frame
in a general curved spacetime. This solution is exact in the test-field
approximation or in the case of a shear-free spacetime without any
other matter. }

\subheader{LMU-ASC-, YITP-15- }

\maketitle

\section{Introduction and Discussion }

The origin of Dark Matter (DM) is one of the main problems in modern
physics. Currently even basic macroscopic properties of DM are not
properly understood, because the only known manifestations of DM are
always through gravity. This fact, together with a more recent discovery
of Dark Energy which also only reveals itself gravitationally, suggest
to look for a suitable modification of General Relativity (GR) on
large scales. The simplest modifications of GR extend it to scalar-tensor
theories. 

A particularly interesting scalar-tensor theory was recently introduced
in \cite{Chamseddine:2013kea} under the name \emph{Mimetic Dark Matter}.
This theory is Weyl-invariant and can naturally describe fluid-like
DM. Indeed, it was realized that \emph{Mimetic Dark Matter} is equivalent
to the fluid-like irrotational dust \cite{Golovnev:2013jxa,Barvinsky:2013mea}
with the \emph{mimetic scalar field} playing the role of the velocity
potential. Moreover, as it was demonstrated in \cite{Chamseddine:2014vna}
a higher-derivative extension of this theory can describe DM with
a small sound speed. The latter property can be very useful to describe
short-scale phenomenology of DM \cite{Capela:2014xta}. This DM corresponds
to a nonstandard imperfect fluid \cite{Mirzagholi:2014ifa}. In the
latter work \cite{Mirzagholi:2014ifa} there was also proposed a mechanism
to generate the proper DM abundance during the radiation dominating
époque. 

Models similar\emph{ Mimetic Dark Matter} to also appear in the IR
limit of the projectable version of Ho\v{r}ava-Lifshitz gravity \cite{Horava:2009uw,Mukohyama:2009mz,Blas:2009yd}
and correspond to a scalar version of the so-called Einstein Aether
\cite{Jacobson:2000xp}. Surprisingly these models can also emerge
in the non-commutative geometry \cite{Chamseddine:2014nxa}.

The origin of \emph{Mimetic Dark Matter }is rooted in a substitution
which is a singular \cite{Deruelle:2014zza} disformal transformation
\cite{Bekenstein:1992pj}. 

In this short paper we discuss in details how this higher-derivative
scalar-tensor theory goes beyond the construction by Horndeski \cite{Horndeski,Deffayet:2011gz,Kobayashi:2011nu},
keeping only one scalar degree of freedom on top of two standard graviton
polarizations. In particular, we consider representations of the theory
in different sets of Weyl-invariant variables \eqref{eq:Invariants_Substitution}
and connect this framework to the singular Brans-Dicke theory \eqref{eq:Beyond_Horndeski}.
Further, in the section \ref{sec:Fluid-picture} we find solution
\eqref{eq:rho_Integral} of equations of motion for the mimetic gravity
in the synchronous reference frame in a general curved spacetime.
This solution is exact in the test-field approximation. Moreover,
this solution is exact in the case of a shear-free spacetime which
is curved by the \emph{Mimetic Dark Matter} only - without any other
external matter.

\section{Different faces of Mimetic Gravity\label{sec:Main-setup} }

Following \cite{Chamseddine:2013kea} let us consider a scalar-tensor
theory with dynamical variables $h_{\mu\nu}$ and $\varphi$ whose
dynamics is given by the action%
\footnote{Throughout the paper we use use the signature convention $\left(+,-,-,-\right)$
and the reduced Planck units where $8\pi G_{\text{N}}=1$. %
} 

\begin{equation}
S_{0}\left[h,\varphi\right]=-\frac{1}{2}\int\mbox{d}^{4}x\,\sqrt{-g\left(h,\varphi\right)}R\left(g\left(h,\varphi\right)\right)\,,\label{eq:Mimetic_EH}
\end{equation}
where the metric $g_{\mu\nu}\left(h,\varphi\right)$ is composed out
of the scalar field $\varphi$ and the metric $h_{\mu\nu}$ in the
following way 
\begin{equation}
g_{\mu\nu}\left(h,\varphi\right)=h^{\alpha\beta}\varphi_{,\alpha}\varphi_{,\beta}\cdot h_{\mu\nu}\,,\label{eq:mimetic_substitution}
\end{equation}
with $\varphi_{,\alpha}=\partial_{\alpha}\varphi$ and $ $$R\left(g\left(h,\varphi\right)\right)$
corresponds to the Ricci scalar calculated for $g_{\mu\nu}$. Thus
$g_{\mu\nu}$ is obtained by a conformal transformation of $h_{\mu\nu}$
with a conformal factor $h^{\alpha\beta}\varphi_{,\alpha}\varphi_{,\beta}$.
By construction the scalar field satisfies the constraint 
\begin{equation}
g^{\mu\nu}\partial_{\mu}\varphi\partial_{\nu}\varphi=1\,,\label{eq:constaraint}
\end{equation}
which is nothing else as the relativistic Hamilton-Jacobi equation
for the particle of the unit mass moving in the spacetime with the
metric $g_{\mu\nu}$.

The metric $g_{\mu\nu}$ is manifestly invariant with respect to the
Weyl transformations of the metric $h_{\mu\nu}$ 
\[
h_{\mu\nu}\rightarrow\Omega^{2}\left(x\right)h_{\mu\nu}\,.
\]
Hence the resulting induced scalar-tensor theory \eqref{eq:Mimetic_EH}
obtained by the substitution \eqref{eq:mimetic_substitution} is manifestly
Weyl-invariant. Making the conformal transformation \eqref{eq:mimetic_substitution}
we obtain the explicit form of the action \eqref{eq:Mimetic_EH} 
\begin{equation}
S_{0}\left[h,\varphi\right]=-\int\mbox{d}^{4}x\,\sqrt{-h}\left(X\, R\left(h\right)+\frac{3}{2}\cdot\frac{h^{\alpha\beta}X_{,\alpha}X_{,\beta}}{X}\right)\,,\label{eq:Beyond_Horndeski}
\end{equation}
where as usual we use the following notation for the standard kinetic
term  
\begin{equation}
X=\frac{1}{2}h^{\alpha\beta}\varphi_{,\alpha}\varphi_{,\beta}\,,\label{eq:X}
\end{equation}
and where we omitted a total derivative. Written as functional on
$X$ this action above \eqref{eq:Beyond_Horndeski} describes the
singular Brans-Dicke theory with the parameter $\omega=-3/2$ %
\footnote{We are thankful to Gilles Esposito-Farese for pointing out this correspondence.%
}. Thus one can obtain \emph{mimetic gravity} or irrotational fluid-like
DM by substituting the kinetic term instead of the scalar field into
the singular / conformal Brans-Dicke theory. Thus there are two different
types of substitutions one can make in order get the same \emph{mimetic
gravity.} Either one substitutes \eqref{eq:mimetic_substitution}
into the Einstein-Hilbert action \eqref{eq:Mimetic_EH} or one substitutes
\eqref{eq:X} into the singular Brans-Dicke theory \eqref{eq:Beyond_Horndeski}.
In the presence of external matter fields a cautionary remark is necessary:
the substitution of the metric $g_{\mu\nu}\left(h,\varphi\right)$
should be performed in the matter part of the action as well. Therefore,
in the singular Brans-Dicke theory the matter should be coupled not
to $h_{\mu\nu}$ but to $2X\cdot h_{\mu\nu}$ before the substitution
of the kinetic term for $\varphi$ instead of the scalar field $X$. 

The action \eqref{eq:Beyond_Horndeski} contains the square of the
second derivative $\varphi_{;\alpha;\beta}$ and goes beyond the usual
scalar-tensor theories of the Horndeski type \cite{Horndeski,Deffayet:2011gz,Kobayashi:2011nu}.
This scalar-tensor theory goes beyond the construction introduced
by Horndeski, but still has only two graviton polarizations plus a
scalar degree of freedom. There is no contradiction there, as the
equations of motion are of the third order for $h_{\mu\nu}$ and forth
order for $\varphi$. However, the theory under consideration is a
gauge theory with the Weyl symmetry, so that physical Weyl-invariant
degrees of freedom satisfy second order equations of motion. Clearly
this way to go beyond the Horndeski scalar-tensor theories is different
from the procedure recently introduced in \cite{Gleyzes:2014dya,Gleyzes:2014qga}. 

It is convenient to get rid of these second derivatives in the action
by promoting $X$ to be a dynamical variable and use a Lagrange multiplier
$\lambda$ in an equivalent action 
\begin{equation}
S_{0}\left[h,\varphi,X,\lambda\right]=-\int\mbox{d}^{4}x\,\sqrt{-h}\left[X\, R\left(h\right)+\frac{3}{2}\cdot\frac{h^{\alpha\beta}X_{,\alpha}X_{\beta}}{X}+\lambda\left(X-\frac{1}{2}h^{\alpha\beta}\varphi_{,\alpha}\varphi_{,\beta}\right)\right]\,.\label{eq:XAction}
\end{equation}
This action is conformally invariant with the corresponding transformations
\begin{align}
 & h_{\mu\nu}\rightarrow\Omega^{2}\left(x\right)h_{\mu\nu}\,,\label{eq:Conformal_transform_weights}\\
 & \varphi\rightarrow\varphi\,,\nonumber \\
 & X\rightarrow\Omega^{-2}\left(x\right)X\,,\nonumber \\
 & \lambda\rightarrow\Omega^{-2}\left(x\right)\lambda\,,\nonumber 
\end{align}
 so that $\lambda$ and $X$ have conformal weight two. 

Now one can make a substitution 
\begin{align}
 & h_{\mu\nu}=\left(2X\right)^{-1}g_{\mu\nu}\,,\label{eq:Invariants_Substitution}\\
 & \varphi=\varphi\,,\nonumber \\
 & X=X\,,\nonumber \\
 & \lambda=2X\,\rho\,,\nonumber 
\end{align}
in the last action \eqref{eq:XAction} in order to rewrite it in terms
of gauge-invariant (Weyl-invariant) variables: $\varphi$, $g_{\mu\nu}=2X\cdot h_{\mu\nu}$
and $\rho=\lambda/\left(2X\right)$. In this way we obtain 
\begin{equation}
S_{0}\left[g,\varphi,\rho\right]=\int\mbox{d}^{4}x\,\sqrt{-g}\left[-\frac{1}{2}R\left(g\right)+\frac{\rho}{2}\left(g^{\alpha\beta}\varphi_{,\alpha}\varphi_{,\beta}-1\right)\right]\,.\label{eq:dust_action}
\end{equation}
This substitution corresponds to a gauge fixing with $X=1/2$ so that
in the original $X$ the required conformal transformation is 
\begin{equation}
\Omega_{X}^{2}=2X\,.\label{eq:fixinig}
\end{equation}
As a result of the substitution \eqref{eq:Invariants_Substitution}
the matter is minimally coupled to the metric $g_{\mu\nu}$. 

A similar discussion on the equivalence of \emph{mimetic} gravity
\eqref{eq:Mimetic_EH} with \eqref{eq:dust_action} was presented
before in \cite{Barvinsky:2013mea}, see also \cite{Golovnev:2013jxa}.
The obtained action \eqref{eq:dust_action} describes irrotational
fluid-like dust with the energy density given by the Lagrange multiplier
$\rho$ and velocity potential $\varphi$ %
\footnote{For the Lagrangian description of dust with vorticity see, \cite{Brown:1992kc,Brown:1994py}.%
}. Here we assume the standard minimal coupling of matter to general
relativity (GR), in particular, matter is not directly coupled%
\footnote{This condition can be violated, if one applies the \emph{mimetic}
ansatz \eqref{eq:mimetic_substitution} in $f\left(R\right)$ theories
\cite{Nojiri:2014zqa}. %
} to the \emph{mimetic} field $\varphi$. 

Interestingly there is an alternative set of gauge-invariant (Weyl-invariant)
variables: $\varphi$, $\hat{g}_{\mu\nu}=\lambda\cdot h_{\mu\nu}$
and $\chi=X/\lambda$. We can obtain an equivalent theory by making
the following ansatz 
\begin{align}
 & h_{\mu\nu}=\lambda^{-1}\hat{g}_{\mu\nu}\,,\label{eq:Invariants_Substitution_Lamba}\\
 & \varphi=\varphi\,,\nonumber \\
 & X=\lambda\,\chi\,,\nonumber \\
 & \lambda=\lambda\,,\nonumber 
\end{align}
into the action \eqref{eq:XAction}. The resulting theory is given
by 
\begin{equation}
S_{0}\left[\hat{g},\varphi,\chi\right]=\int\mbox{d}^{4}x\,\sqrt{-\hat{g}}\left[\frac{1}{2}\hat{g}^{\alpha\beta}\varphi_{,\alpha}\varphi_{,\beta}-\left(\chi\, R\left(\hat{g}\right)+\frac{3}{2}\cdot\frac{\hat{g}^{\alpha\beta}\chi_{,\alpha}\chi_{\beta}}{\chi}+\chi\right)\right]\,,\label{eq:Unity_Lambda_Action}
\end{equation}
which corresponds to the a gauge fixing with $\lambda=1$ so that
in terms of the original $\lambda$ the required conformal transformation
is 
\begin{equation}
\Omega_{\lambda}^{2}=\lambda\,.\label{eq:fixinig-1}
\end{equation}
Here, contrary to the previous case, the matter except of $\varphi$
is still coupled not to $\hat{g}_{\mu\nu}$ but to $2\chi\hat{g}_{\mu\nu}$.
The action $ $$S_{0}\left[\hat{g},\varphi,\chi\right]$ seems to
be a novel description of the irrotational fluid-like dust. 

Further in this paper we will concentrate on the imperfect DM from
\cite{Chamseddine:2014vna,Mirzagholi:2014ifa,Capela:2014xta}. 

The action in the convenient invariant variables is 
\begin{equation}
S_{\gamma}\left[g,\rho,\varphi\right]=\int\mbox{d}^{4}x\,\sqrt{-g}\left(-\frac{1}{2}R\left(g\right)+\frac{\rho}{2}\left(g^{\mu\nu}\partial_{\mu}\varphi\partial_{\nu}\varphi-1\right)+\frac{1}{2}\gamma\left(\Box_{g}\varphi\right)^{2}\right)\,,\label{eq:action}
\end{equation}
where $\rho$ is a Lagrange multiplier field, $\varphi$ is the \emph{mimetic}
field, $\gamma$ is a free parameter%
\footnote{Contrary to \cite{Mirzagholi:2014ifa}, in this work we will only
consider $\gamma=const$ neglecting the short transition period in
the early universe where $\gamma\left(\varphi\right)$. %
} and $\Box_{g}=g^{\mu\nu}\nabla_{\mu}\nabla_{\nu}$ with $\nabla_{\mu}\left(\,\,\right)=\left(\,\,\right)_{;\mu}$
being the covariant derivative. This system was introduced in \cite{Chamseddine:2014vna}
and it generalizes \cite{Lim:2010yk,Chamseddine:2013kea}, see also
\cite{Haghani:2014ita}. It is interesting to formulate this theory
as hinger-derivative scalar-tensor theory with Weyl symmetry. This
is can be accomplished via the \emph{mimetic} ansatz \eqref{eq:mimetic_substitution}
into 
\[
S_{\gamma}\left[g,\varphi\right]=\int\mbox{d}^{4}x\,\sqrt{-g}\left(-\frac{1}{2}R\left(g\right)+\frac{1}{2}\gamma\left(\Box_{g}\varphi\right)^{2}\right)\,,
\]
with the resulting action 
\[
S_{0}\left[h,\varphi\right]=-\int\mbox{d}^{4}x\,\sqrt{-h}\left[X\, R\left(h\right)+\frac{3}{2}\cdot\frac{h^{\alpha\beta}X_{,\alpha}X_{,\beta}}{X}-\frac{1}{2}\gamma\left(\Box_{h}\varphi+\frac{h^{\alpha\beta}\varphi_{,\alpha}X_{\beta}}{X}\right)^{2}\right]\,,
\]
where as before we denoted $X=\tfrac{1}{2}h^{\alpha\beta}\varphi_{,\alpha}\varphi_{,\beta}$.
This is another example of a higher-derivative scalar-tensor theory
beyond Horndeski, but with just two graviton polarizations and one
scalar degree of freedom \cite{Chamseddine:2014vna}. 

An equivalent action \cite{Mirzagholi:2014ifa} to describe the dynamics
of the system without explicit higher derivatives is 
\[
S_{\gamma}\left[g,\rho,\varphi,\theta\right]=\int\mbox{d}^{4}x\,\sqrt{-g}\left[-\frac{1}{2}R\left(g\right)+\frac{\rho}{2}\left(\varphi^{,\mu}\varphi_{,\mu}-1\right)-\gamma\left(\varphi_{,\mu}\theta^{,\mu}+\frac{1}{2}\theta^{2}\right)\right]\,,
\]
where $\theta$ is an auxiliary field. The equation of motion for
this auxiliary field gives 
\begin{equation}
\theta=\Box_{g}\varphi=\frac{1}{\sqrt{-g}}\partial_{\mu}\left(\sqrt{-g}\, g^{\mu\nu}\partial_{\nu}\varphi\right)\,.\label{eq:teta}
\end{equation}
The theory is shift-invariant: symmetric with respect to $\varphi\rightarrow\varphi+c$,
thus there is a conserved Noether current which is given by 
\begin{equation}
J_{\mu}=\rho\varphi_{,\mu}-\gamma\theta_{,\mu}\,.\label{eq:Noether_General}
\end{equation}
 So that the equation of motion for $\varphi$ is corresponds to the
conservation of the Noether current 

\begin{equation}
\nabla_{\mu}J^{\mu}=0\,.\label{eq:Phi_EOM}
\end{equation}
This equation can be written in the form 
\begin{equation}
g^{\mu\nu}\partial_{\mu}\varphi\,\partial_{\nu}\rho+\theta\rho=\gamma\Box_{g}\theta\,.\label{eq:expanded_eom}
\end{equation}
The energy-momentum tensor (EMT) is
\begin{align}
 & T_{\mu\nu}=\frac{2}{\sqrt{-g}}\frac{\delta S_{\gamma m}}{\delta g^{\mu\nu}}=\rho\,\varphi_{,\mu}\varphi_{,\nu}+\gamma\left[g_{\mu\nu}\left(\varphi_{,\alpha}\theta^{,\alpha}+\frac{1}{2}\theta^{2}\right)-\varphi_{,\mu}\theta_{,\nu}-\varphi_{,\nu}\theta_{,\mu}\right]\,,\label{eq:EMT_General}
\end{align}
 where we subtract from the action the Einstein-Hilbert part: $S_{\gamma m}=S_{\gamma}-S_{\text{EH}}$,
and we have assumed that the constraint \eqref{eq:constaraint} is
satisfied together with the equations of motion.

\section{Solution in Synchronous Frame \label{sec:Fluid-picture} }

Because of the constraint \eqref{eq:constaraint} the four velocity
$u_{\mu}=\partial_{\mu}\varphi$ is tangential to the time-like geodesics
\cite{Lim:2010yk} 
\begin{equation}
a^{\mu}=u^{\lambda}\nabla_{\lambda}u^{\mu}=\nabla^{\lambda}\varphi\nabla_{\lambda}\nabla^{\mu}\varphi=\frac{1}{2}\nabla^{\mu}\left(\nabla^{\lambda}\varphi\nabla_{\lambda}\varphi\right)=0\,.
\end{equation}
Therefore we will call the $\partial_{\mu}\varphi$ frame \textendash{}
the natural or the geodesic frame. It is convenient to consider local
rest frame (LRF) given by a natural choice $u_{\mu}=\partial_{\mu}\varphi$
and go the following synchronous frame
\begin{align}
 & \varphi=\tau\,,\label{eq:Synchronous frame}\\
 & ds^{2}=d\tau^{2}-\ell_{ik}\left(\tau,\mathbf{x}\right)dx^{i}dx^{k}\,.\nonumber 
\end{align}
In this frame for every scalar quantity $\mathcal{O}$ 
\[
\dot{O}=u^{\mu}\partial_{\mu}O=\partial_{\tau}O\,.
\]
The auxiliary field $\theta$ plays the role of expansion $\theta=\nabla_{\mu}u^{\mu}$
and can be written as 
\[
\theta=\partial_{\tau}\ln\sqrt{\ell}\,.
\]
The equation of motion \eqref{eq:expanded_eom} takes the form of
the ordinary differential equation 
\[
\partial_{\tau}\rho+\partial_{\tau}\ln\sqrt{\ell}\,\rho=\gamma\,\frac{1}{\sqrt{\ell}}\partial_{\mu}\left(\sqrt{\ell}\, g^{\mu\nu}\partial_{\nu}\partial_{\tau}\ln\sqrt{\ell}\right)\,.
\]
For $\gamma=0$ this equation can be integrated as 
\[
\rho_{0}\left(\tau,\mathbf{x}\right)=\frac{Q\left(\mathbf{x}\right)}{\sqrt{\ell}}\,,
\]
where $Q\left(\mathbf{x}\right)$ is a free function. For nonvanishing
$\gamma$ one can use the method of variation of the constant and
substitute 
\[
\rho\left(\tau,\mathbf{x}\right)=\frac{Q\left(\tau,\mathbf{x}\right)}{\sqrt{\ell}}\,,
\]
to obtain 
\[
\partial_{\tau}Q=\gamma\,\partial_{\mu}\left(\sqrt{\ell}\, g^{\mu\nu}\partial_{\nu}\partial_{\tau}\ln\sqrt{\ell}\right)\,,
\]
so that 
\[
Q\left(\tau,\mathbf{x}\right)=\varrho_{0}\left(\mathbf{x}\right)+\gamma\,\sqrt{\ell}\,\partial_{\tau}\partial_{\tau}\ln\sqrt{\ell}-\gamma\partial_{i}\int_{0}^{\tau}d\tau'\sqrt{\ell}\ell^{ik}\partial_{k}\partial_{\tau}\ln\sqrt{\ell}\,,
\]
where $\varrho_{0}\left(\mathbf{x}\right)$ is a free function. The
corresponding solution is 
\begin{equation}
\rho\left(\tau,\mathbf{x}\right)=\frac{\varrho_{0}\left(\mathbf{x}\right)}{\sqrt{\ell}}+\gamma\left(\partial_{\tau}\partial_{\tau}\ln\sqrt{\ell}-\frac{1}{\sqrt{\ell}}\partial_{i}\int_{0}^{\tau}d\tau'\sqrt{\ell}\ell^{ik}\partial_{k}\partial_{\tau'}\ln\sqrt{\ell}\right)\,.\label{eq:rho_Integral}
\end{equation}
Only the first term in this formula corresponds to the standard DM.
In the test-field approximation $\ell$ does not depend on $\rho$
through the Einstein equations and this formula above is sufficient
to find the evolution of $\rho$. 

Now let us see what happens, if we do take into account how \emph{mimetic}
matter curves the spacetime. 

First we need to recall some details from the hydrodynamical picture
from \cite{Mirzagholi:2014ifa}. The shift-symmetry Noether current
can be decomposed using the local rest frame (LRF) given by a natural
choice $u_{\mu}=\partial_{\mu}\varphi$ as 
\begin{equation}
J_{\mu}=nu_{\mu}-\gamma\bot_{\mu}^{\lambda}\nabla_{\lambda}\theta\,,\label{eq:Noether decomposed}
\end{equation}
where the shift-charge density is 
\begin{equation}
n=J^{\mu}u_{\mu}=\rho-\gamma\partial_{\tau}\theta\,,\label{eq:Charge density}
\end{equation}
and as usual we use the following notation for the projector to the
hypersurface orthogonal to $u^{\mu}$: 
\begin{equation}
\bot_{\mu\nu}=g_{\mu\nu}-u_{\mu}u_{\nu}\,.\label{eq:projector}
\end{equation}
 Hence for the charge density we obtain 
\begin{equation}
n\left(\tau,\mathbf{x}\right)=\frac{\varrho_{0}\left(\mathbf{x}\right)}{\sqrt{\ell}}-\frac{\gamma}{\sqrt{\ell}}\partial_{i}\int_{0}^{\tau}d\tau'\sqrt{\ell}\ell^{ik}\partial_{k}\theta\,.\label{eq:charge_density_solution}
\end{equation}
Decomposition of the EMT \eqref{eq:EMT_General} reads 
\begin{equation}
T_{\mu\nu}=\varepsilon u_{\mu}u_{\nu}-p\perp_{\mu\nu}+q_{\mu}u_{\nu}+q_{\nu}u_{\mu}\,,\label{eq:EMT fluid}
\end{equation}
where the energy density in the LRF given by 
\begin{equation}
\varepsilon=T_{\mu\nu}u^{\mu}u^{\nu}=\rho-\gamma\left(\dot{\theta}-\frac{1}{2}\theta^{2}\right)\,,\label{eq:energy density}
\end{equation}
while the pressure is 
\begin{equation}
p=-\frac{1}{3}T^{\mu\nu}\perp_{\mu\nu}=-\gamma\left(\dot{\theta}+\frac{1}{2}\theta^{2}\right)\,,\label{eq:Pressure}
\end{equation}
and the energy flux 
\begin{equation}
q_{\mu}=\bot_{\mu\lambda}T_{\sigma}^{\lambda}u^{\sigma}=-\gamma\bot_{\mu}^{\lambda}\nabla_{\lambda}\theta=-\gamma\delta_{\mu}^{i}\partial_{i}\theta\,,\label{eq:energy flux}
\end{equation}
so that 
\begin{equation}
T_{i}^{0}=-\gamma\partial_{i}\theta\,.\label{eq:T0i}
\end{equation}
Further we will need the Raychaudhuri equation for the timelike geodesics
\begin{equation}
\dot{\theta}=-\frac{1}{3}\theta^{2}-\sigma_{\mu\nu}\sigma^{\mu\nu}-R_{\mu\nu}u^{\mu}u^{\nu}\,,\label{eq:Raychaudhuri}
\end{equation}
where $\sigma_{\mu\nu}$ is the shear tensor
\[
\sigma_{\mu\nu}=\frac{1}{2}\left(\bot_{\mu}^{\lambda}\nabla_{\lambda}u_{\nu}+\bot_{\nu}^{\lambda}\nabla_{\lambda}u_{\mu}\right)-\frac{1}{3}\bot_{\mu\nu}\theta\,,
\]
$\sigma^{2}=\sigma_{\mu\nu}\sigma^{\mu\nu}$. Now we are equipped
to look at the Einstein equations in the synchronous frame, see \cite{Landau:1982dva}.
Here we will only need $R_{i}^{0}-$equation 
\[
R_{i}^{0}=\frac{1}{2}\left(\varkappa_{i;k}^{k}-\varkappa_{k;i}^{k}\right)=8\pi G_{\text{N}}\left(T_{i}^{0}+\mathcal{T}_{i}^{0}\right)\,,
\]
where $\mathcal{T}_{i}^{0}$ is the energy momentum for all other
matter fields and 
\[
\varkappa_{ik}=\frac{\partial\ell_{ik}}{\partial\tau}\,.
\]
After simple algebra one obtains 
\[
\varkappa_{ik}=2\left(-\sigma_{ik}+\frac{1}{3}\ell_{ik}\theta\right)\,,
\]
so that using \eqref{eq:T0i} we can rewrite the $0i$ Einstein equation
as 
\[
\theta_{,i}=-\frac{12\pi G_{\text{N}}}{1-12\pi G_{\text{N}}\gamma}\mathcal{T}_{i}^{0}-\frac{3}{2}\frac{\sigma_{i;k}^{k}}{1-12\pi G_{\text{N}}\gamma}\,.
\]
Thus if there is no shear and no external matter the integral in \eqref{eq:rho_Integral}
and in \eqref{eq:charge_density_solution} vanishes and the behavior
of the mimetic imperfect DM is only different from a fluid-like dust
by the derivative of the expansion scalar $\theta$. It is interesting
to note that this $0i$ Einstein equation above suggests that there
is an effective Newton constant 
\[
G_{\text{eff}}=\frac{G_{\text{N}}}{1-12\pi G_{\text{N}}\gamma}\,,
\]
which was found in the cosmological context in \cite{Mirzagholi:2014ifa}.

\section*{Acknowledgements\label{sec:Acknowledgements}}

\addcontentsline{toc}{section}{Acknowledgements} 

It is a pleasure to thank Eugeny Babichev, Cedric Deffayet, Gilles
Esposito-Farese, Andrei Frolov, Viatcheslav Mukhanov, Shinji Mukohyama,
Sabir Ramazanov, Ignacy Sawicki, Sergey Sibiryakov for very useful
discussions and criticisms. AV is thankful to the Laboratoire de Gravitation
et Cosmologie Relativistes ($\mathcal{G}\mathbb{R}\varepsilon\mathbb{C}\mathcal{O}$),
Universiti Pierre et Marie Curie (Paris VI) (UPMC) and Institut d'Astrophysique
de Paris (IAP) for a very warm hospitality during the final stages
of this project. The work of AV was supported by the J. E. Purkyn\v{e}
Fellowship of the Czech Academy of Sciences and by the Grant Agency
of the Czech Republic under the grant P201/12/G028. 

\bibliographystyle{utcaps}
\addcontentsline{toc}{section}{\refname}\bibliography{Mimetic}

\end{document}